\documentstyle[twocolumn,epsf,aps]{revtex}
\def\centerplot#1#2{
\begin{center}
\leavevmode
{\epsfxsize=#2\textwidth \epsfbox{#1}}
\end{center}
}
\begin{document}
\draft
\title{
Spin glass behavior of frustrated 2-D Penrose lattice
in the classical planar model }
\author{R.W. Reid, S.K. Bose, and  B. Mitrovi\'{c}}                     
\address{
Physics Department, Brock University, St. Catharines, Ontario
L2S 3A1 CANADA}
\date{Received 29 December 1995}
\twocolumn[
\maketitle \hsize\textwidth \leftskip=0.10753\textwidth \rightskip\leftskip
\begin{center}
\vspace{-0.4in}
\begin{minipage}{6in}
\begin{abstract}
Via extensive Monte Carlo studies we show that the frustrated XY Hamiltonian
on a 2-D Penrose lattice admits of a spin glass phase at low temperature. 
Studies of the Edwards-Anderson order
parameter, spin glass susceptibility, and local (linear) susceptibility point
unequivocally to a paramagnetic to spin glass transition as the temperature is
lowered. 
Specific heat shows a rounded peak at a temperature above the spin glass
transition temperature, as is commonly observed in spin glasses.  Our results
strongly suggest  that the critical point exponents are the same as obtained by Bhatt and
Young in the ${\pm}J$ XY model on a square lattice. 
However, unlike in the latter case,  the critical temperature is
clearly finite (nonzero). The results imply that a quasiperiodic 2-D array of
superconducting grains in a suitably chosen transverse magnetic field should
behave as a superconducting glass at low temperature.
\end{abstract}
\end{minipage}
\end{center}
]
The notion of frustration, first introduced by Toulouse$^1$, has played a
key role in the theoretical understanding of spin glasses. It is now believed
that frustration is a necessary condition for the existence of a spin glass
phase. However, all real systems (alloys) exhibiting spin glass behavior have an
additional common feature, namely the topological disorder of  magnetic
component(s). On the theoretical front, models of uniformly frustrated spin
systems without disorder have often failed to show spin glass behavior, and
the debate as to whether disorder, in addition to frustration, is necessary
in a spin glass continues. In this paper we present strong evidence that a
frustrated  XY Hamiltonian on a 2-D quasiperiodic lattice admits of a low
temperature spin glass phase. The degree of frustration in this case varies
widely from one lattice point to another, but in a predictable way determined
by the quasiperiodic structure.

We consider the Hamiltonian for the XY model describing the interaction between
2-D spin vectors with orientations $\theta_i$ and $\theta_j$ situated at
lattice sites $i$ and $j$ via a nearest neighbor coupling parameter $J$:
\begin{equation}
H = -J\sum_{[ij]}\cos(\theta_{i}-\theta_{j}+A_{ij}),
\end{equation}
where the summation is restricted to nearest neighbor pairs   $[ij]$. 
The parameter $A_{ij}$ controls the frustration in the model. In the context
of an array of superconducting grains, $\theta_i$ is the phase of the grain
$i$ and the above Hamiltonian can be seen as describing the resulting Josephson junction of the grains ``minimally coupled''
to a transverse magnetic field with vector potential ${\bf A}$ with
\begin{equation}
A_{ij} = \frac{{2\pi}}{\Phi_0}\int^{\bf r_j}_{\bf r_i}{\bf A}\cdot d{\bf l}\:.
\end{equation}
$\Phi_0$ is the elementary flux quantum ${\frac{hc}{2e}}$ associated with the Cooper pairs, and ${\bf r_i}$ denotes the lattice sites.
Here the magnetic field acts as the source of frustration: an $A_{ij}$ which is
an odd multiple of $\pi$ essentially renders the bond $[ij]$ negative. 
 
The directed sum of $A_{ij}$ about a plaquette in a 2-D lattice can be written
as 2$\pi f$, where $f$ is the flux through the plaquette in units of $\Phi_0$. 
The 2-D Penrose lattice$^2$ is composed of two (fat and thin) rhombic unit
cells(plaquettes).  Since the ratio of the areas of the fat and the thin
rhombuses in the Penrose lattice is the Golden Mean ($\tau$), only one set
of plaquettes can be fully frustrated at a time  with a suitable choice of the magnetic field giving $f=1/2$. The flux $f$ through the individual plaquettes in
the other set will then be an irrational number.

In what follows we will present
results for the case where the `thin rhombus' is fully frustrated, mentioning
at the outset that the results
for the fully frustrated `fat rhombus' case are qualitatively similar.
All our results are obtained via Monte Carlo (MC) simulation based on the
Metropolis
algorithm$^3$ using periodic boundary conditions on periodic Penrose lattices.  
We have cooled our systems in a quasi-static manner, starting from a high
temperature $(T$(in units of $J$)$>2.0)$ random configuration and then heated
the system in the
same quasi-static fashion. Since we performed the simulation in $n$ blocks,
the heating and cooling data are obtained by averaging over these blocks,
with the error bars representing the standard deviation,
obtained by dividing the square root of sum of squares of the deviations from
the mean by $\sqrt{n-1}$, instead of $\sqrt{n}$.
We then perform a `grand average' over the heating and cooling data.

Magnetic moment per lattice point calculated for periodic
Penrose lattices of various sizes is found to be small ($<0.02$)
over the entire temperature range. The magnitude of the moment decreases
steadily with the size of the cluster, suggesting that the magnetization
is strictly zero in the thermodynamic limit.
To study the freezing of the spins at the lattice sites we calculate     
the Edwards-Anderson$^4$ order parameter, a measure of breakdown
of ergodicity in the system. In Fig.1 we show this order parameter defined by 
\begin{equation}
q_{EA} = \sum_{i}\langle \vec S_{i} \rangle ^2 ,
\end{equation}
where ${\vec S_i}$ = $(\cos\theta_i,\sin\theta_i)$,
and ${\langle \hskip .5pc\relax \rangle}$ denotes a 
canonical ensemble average. In a completely frozen system $q_{EA}$ is unity,
while for a completely ergodic system it is zero.
\begin{figure*}[htbp]
\centerplot{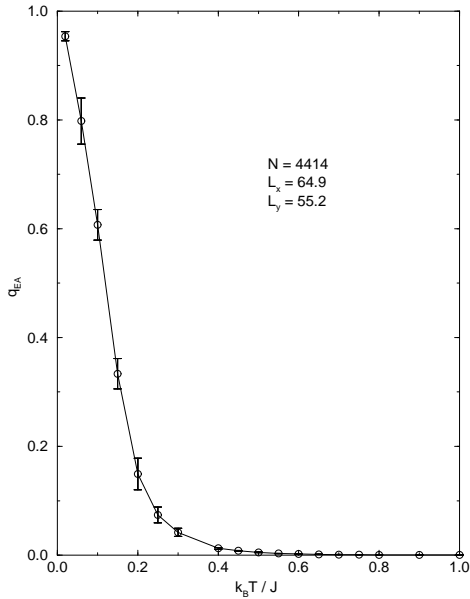}{0.4}
\vspace{-0.25in}
\caption{
Edwards-Anderson order parameter as a function of temperature
for N=4414. The linear dimensions of the lattice in $x$ and $y$ directions
are indicated as $L_X$ and $L_y$, respectively}
\label{Fig.1.}
\end{figure*}
This order parameter shows a monotonic decrease with increasing temperature,
clearly vanishing at temperatures beyond 0.5 for the 4,414 site periodic
Penrose lattice shown in Fig.1. 
These results were obtained by averaging
over 5 blocks of 60,000 configurations, generated after equilibrium was
achieved. Since the vanishing of the order parameter with
a long tail is a consequence of the finite system size, it is expected that
the tail region will decrease with increasing system size
and eventually disappear
in the thermodynamic limit. However, we find that this tail persists, even for
our largest system $(N=11,556)$ and, consequently, the
spin glass transition temperature cannot be appropriately determined from
Fig.1. Thus, we use other quantities to provide an estimate of the transition
temperature $T_f$. It is clear, however, that the system has a low temperature
phase with a nonzero order parameter $q_{EA}$.
 
Local (linear) susceptibility per spin, calculated from the fluctuations in the
magnetization (net magnetic moment $m$ for a lattice of $N$ sites),
\begin{equation}
\chi = \frac{\langle m^2 \rangle - \langle m \rangle^2}{Nk_BT},
\end{equation}
is shown in Fig.2.
At low temperatures, $(0.02 - 0.2)$ the results are obtained by averaging 
over 5 blocks of 125,000 MC steps, while 5 blocks of 15,000-45,000 steps
were used for higher temperatures. 
The large hysteresis in the low temperature region indicates a
high number of metastable states, which is a characteristic of spin glasses.
These metastable states give rise to large error bars in
${\langle m^2 \rangle - \langle m \rangle^2}$  at low
temperatures, which are further accentuated by a division by $T$ in Eq.(4).
Although we feel that it might be possible to reduce the size of these
error bars, this would require very long runs and one must also ensure that
the system does not become trapped in one of these metastable states.
Nevertheless, despite the large error bars, a cusp-like feature in $\chi$
is clearly visible at $T_f\sim0.15$. We find a saturation in this cusp with
respect to system size, which is consistent with spin glass
behavior. Note that the high temperature (above $T_f$) phase is strictly
paramagnetic. Here the decrease in $\chi$ is
Curie-like, with $T \chi$ becoming a constant as shown in the inset 
of Fig.2.
\begin{figure*}[htbp]
\centerplot{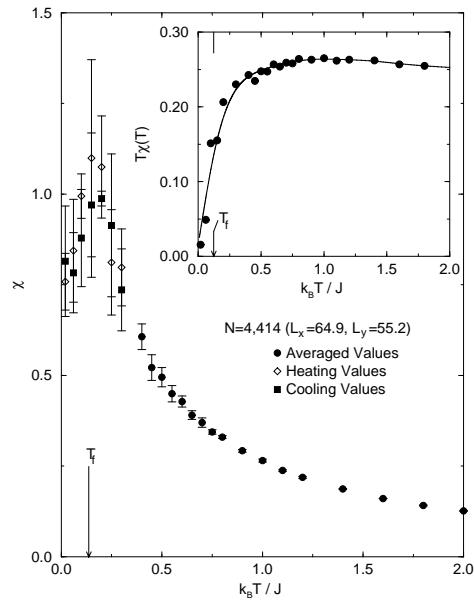}{0.4}
\vspace{-0.25in}
\caption{
Linear susceptibility as a function of temperature,
with a peak around T=0.15.}
\label{Fig.2.}
\end{figure*}

Specific heat obtained from the fluctuations in the energy is shown in Fig.3
for various system sizes. The results are averages between heating and cooling,
with the low temperature results having a somewhat larger hysteresis.
All averages were obtained after equilibriating, however, the high temperature
values were obtained by averaging over 5 blocks of 15,000 steps, whereas 5
blocks of 45,000 steps were used for the low temperatures.
The result for the unfrustrated case is shown in the inset. Both frustrated and
unfrustrated cases show saturation in specific heat with respect to system size.
In the unfrustrated case the peak in the specific heat is at 1.10, which is
beyond the Kosterlitz-Thouless (KT) transition$^5$ temperature $T_{KT}$.
The specific heat
peak for the frustrated case is more rounded relative to the unfrustrated
case and occurs at a temperature lower than $T_{KT}$. This temperature is,
however, higher than the temperature $T_f$ at which, we believe, the
Edwards-Anderson
order parameter goes to zero or a cusp appears in the linear susceptibility.
\begin{figure*}[htbp]
\centerplot{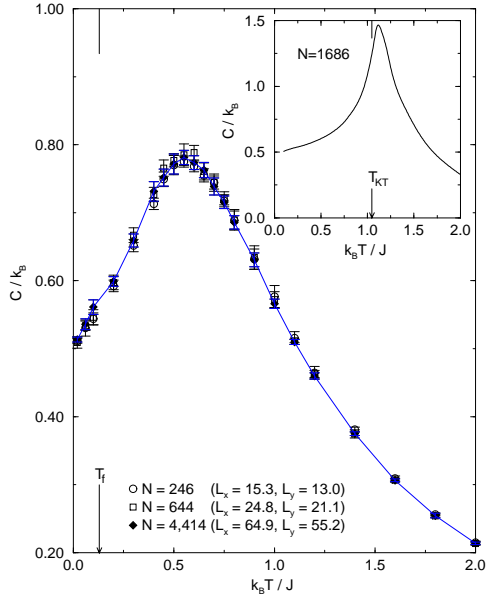}{0.4}
\vspace{-0.25in}
\caption{
Specific heat as a function of temperature.
The inset shows the specific heat for the unfrustrated
case with the peak slightly beyond the KT transition
temperature}
\label{Fig.3.}
\end{figure*}
The saturation in the peak height of the specific heat is a consequence of the
fact that it appears at a temperature at which the spin glass correlation is
finite. Note that in both the unfrustrated and the frustrated cases the zero
temperature specific heat approaches a value of 0.5k$_B$, consistent with a
linear spin-wave theory.
 
In a ferromagnet, the approach to the ferromagnetic phase from temperatures 
above the Curie temperature $T_C$ is accompanied by a dramatic increase in
the range of
the spin correlations, which then diverges at $T_C$. A corresponding   
phenomenon occurs in spin glasses. However, it is not the spin correlation
function $\langle \vec S_i\cdot\vec S_j \rangle$, but rather its square that
acquires a long-range. This leads to the divergence, at the spin glass    
transition temperature $T_f$, of the spin glass susceptibility
\begin{equation}
\chi_{_{SG}} = \frac{1}{N} \sum_{ij}\langle \vec S_i\cdot\vec S_j \rangle^2
\hskip 0.8pc\relax (T > T_f)\:.
\end{equation}
$\chi_{_{SG}}$ satisfies a finite-size scaling relation of the form$^5$
\begin{equation}
\chi_{_{SG}} = L^{2-\eta}\bar\chi(L^{1/\nu}(T-T_f))\:,
\end{equation}
where $\bar\chi$ is the scaling function, $L$ is the system length, $\nu$ is 
the exponent for spin glass correlation length $\xi$ for $T \geq T_f$, and 
$\eta$ describes the power law decay of the spin glass correlation at $T_f$.
For a square lattice $L=\sqrt{N}$, and the above scaling relation can also
be expressed in terms of $N$. We have used this scaling relation, expressed
in terms of $N$, for periodic Penrose lattices of various size to study 
$\chi_{_{SG}}$. To ensure a proper convergence of $\chi_{_{SG}}$, we have
averaged over 5 blocks of 40,000-60,000 steps at low temperatures $(<0.2)$ and
and 5 blocks of 60,000-80,000 steps at higher temperatures. By examining our
results every 5,000 steps, we find little change in $\chi_{_{SG}}$ over the
last 5,000-10,000 steps. Thus, we estimate that these chain lengths produced
at least a 95\% convergence in $\chi_{_{SG}}$.
\begin{figure*}[htbp]
\centerplot{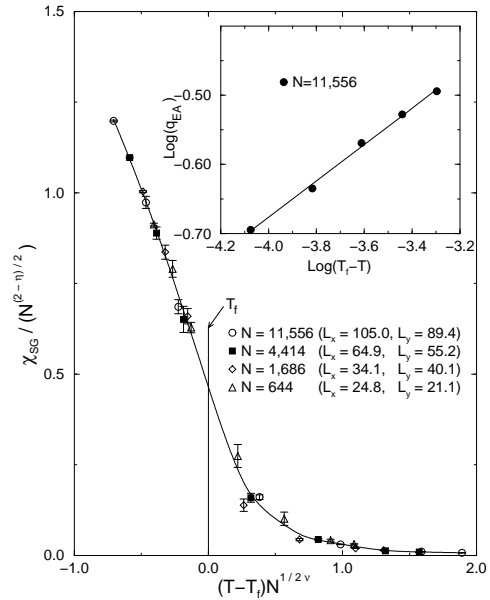}{0.4}
\vspace{-0.25in}
\caption{
Scaling  plot for spin glass susceptibility.
The inset shows the log-log plot of the Edwards-Anderson order
parameter with (T-T$_f$) with T$_f$=0.137 obtained from the fit
to the scaling relation (6) with $\eta$=0.02 and $\nu$=2.6 , as given
by Bhatt and Young$^5$. The slope of this plot yields a value of 0.26
for the exponent $\beta$, in agreement with the hyperscaling
relation $\beta =\nu\eta/2$.}
\label{Fig.4.}
\end{figure*}

In Fig.4, we show the scaling behavior of $\chi_{_{SG}}$ in
terms of $N$ for $T_f=0.137$, where we have used the values of $\nu=2.6$ and
$\eta=0.2$ reported by Bhatt and Young$^6$ in their study of $\pm J$ XY model
on square lattices. These results represent the average over cooling and heating
and lie within the $0.02-0.45$ temperature range. A choice of $T_f=0$ results
in four distinctly separate curves, indicating a clear breakdown of the
scaling relation. If we try to obtain a good fit to the scaling curve with
$T_f=0$, then the exponents $\nu$ and $\eta$ deviate drastically from the values
obtained by Bhatt and Young$^6$ and others (see references in Bhatt and Young),
obtained for 2-D XY spin glass models.

Furthermore, we have examined the Edwards-Anderson parameter in the
vicinity of $T_f$, where the relation $q_{EA}\sim (T-T_f)^\beta$ is supposed
to hold. We find that our value of $T_f=0.137$ yields a value of $\beta=0.26$,
(see inset of Fig.4)
which satisfies the hyperscaling relation $\beta =\frac{\nu \eta}{2}$
valid for our 2-D system. Note that this value of $T_f$ is clearly
consistent with our results for the susceptibility $\chi$, which shows a
cusp at $T\sim0.15$ (Fig.2).
The inescapable conclusion seems to be that the frustrated 2-D
Penrose lattice has a spin glass transition temperature above zero
with exponents that are equal (similar) to those obtained for random XY models
on square lattices.

In summary, we have shown, via the Edwards-Anderson order parameter, spin glass
susceptibility and local (linear) susceptibility, the existence of a 
low temperature spin glass phase in frustrated 2-D Penrose lattice. Our    
results for magnetization and specific heat also support this 
picture. Our detailed study of the unfrustrated case (not reported here)
reveal a behavior similar to that of a square lattice$^7$ with
a somewhat higher KT transition temperature.
We would like to add that similar studies by us performed on other frustrated
quasiperiodic lattices (such as the octagonal lattice) also show a low
temperature spin glass phase.

A few comments are in order at this stage. Our results are consistent with those
of Halsey$^8$, who finds a spin glass phase for the frustrated XY model on a
square lattice with an irrational flux through the plaquettes. Note that we
can fully frustrate only one of the two elementary plaquettes at one time, 
the corresponding flux through the other plaquette being irrational. In
both cases, the model studied by Halsey$^8$ and the frustrated Penrose lattice,
the parameter $\cos(A_{ij})$, which determines the nature of the coupling along
the bond $[ij]$, shows a similar distribution in its values, varying widely
between 1 and -1. This is different from most of the fully frustrated regular
lattices studied so far. For example, the fully frustrated square lattice
studied by Teitel and Jayaprakash$^9$ or the  triangular lattice studied
by Shih and Stroud$^{10}$, where the nature of the transition is seen to change
from  KT to the Ising type, has a much narrower distribution of
$\cos(A_{ij})$.

The experimental implication of our study is that
a quasiperiodic array (such as the Penrose or octagonal lattice)
of Josephson junctions in a suitably chosen transverse magnetic field  
should behave as a superconducting glass at low temperature.
Via advanced microfabrication techniques$^{11}$, it should be possible to
generate such quasiperiodic arrays of superconducting grains. Experimental
investigation on a 2-D fractal (Sierpinski-gasket) network has been
reported$^{12}$. Halsey$^8$ has pointed out that for superconducting
arrays with low normal-state
resistivities the glass transition should basically appear as a mean-field
transition, with fluctuation effects being barely observable. For arrays
with high normal-state resistivities the fluctuation effects will cause
the glass transition to deviate substantially from a mean-field transition,  
with noticeable  system-dependent details.
An important feature of  such glassy superconductors, as pointed out
by Ebner and Stroud$^{13}$, is a large difference between their dc and ac  
susceptibilities.

Finally, a comment about the lower critical dimension for the spin glass
phase. It is widely believed that the lower critical dimension for the
spin glass phase is greater than two$^{14}$. This is corroborated by
Monte Carlo studies by Binder and co-workers$^{15}$ 
(see also ref. 14) and Young and co-workers$^{6}$ on square lattices. Our
results indicate that quasiperiodic 2-D lattices may provide an exception.
The glassy behaviour we are reporting is truly the behavior of macroscopic
systems and not a reflection of some other model masked by the finite
system size.
This behavior is guaranteed for the particular choice of frustration
considered in this paper, dictated by the quasiperiodic lattice structure
and the magnetic field.
Unlike in the above studies$^{6,14,15}$ with completely random 
distributions of the coupling constants along the bonds, the effective coupling
constants in our model are nonrandom.
However, our results obtained for the particular magnetic field considered
are amenable to experimental verification and merit attention.
\vspace{0.2in}

Financial support for this work was provided by  the Natural Sciences and 
Engineering Research Council of Canada.

\begin{center}
ERRATUM
\end{center}
The authors regret that the following errors remained undetected
during the proof stage.\\
(1) In the abstract `${\pm J}\:XY$ model' should read `${\pm J}$ Ising model'.\\
(2) Ref. 6 should read:\\
(a)R.N. Bhatt and A.P. Young, Phys. Rev. B {\bf 37}, 5606 (1988);
(b) S. Jain and A.P. Young, J. Phys. C: Solid State Phys. {\bf 19}, 3913
(1986).\\
(3) In the caption of Fig. 4, Ref. 5 should be changed to Ref. 6(a),
and $\eta =0.02$ should be changed to $\eta=0.2$.\\
(4) Ref.5 immediately preceding Eq.(6) should be changed to Ref. 6(a).\\
(5) Regarding the exponents, note that our results agree much better with
those of Bhatt and Young$^{6(a)}$ (${\pm J}$ Ising model) than with those
of Jain and Young$^{6(b)}$ (${\pm J}\:XY$ model). Accordingly, references to
${\pm J}\:XY$ model study by Bhatt and Young$^6$ should be changed to
${\pm J}$ Ising model study by Bhatt and Young$^{6(a)}$ at appropriate places
in the main body of the paper.

One final comment: It is not clear to us at this stage whether the agreement
with the ${\pm J}$ Ising model is purely coincidental, or there is a connection
with the result of S. Teitel and C. Jayaprakash, Phys. Rev. B {\bf 27}, 598
(1983) (Ref. 9) on square lattice, where the nature of the transition is seen
to change from KT to the Ising type as a result of frustration.

The authors are thankful to A.P. Young for pointing out the referencing error
regarding the ${\pm J}\:XY$ and ${\pm J}$ Ising model calculations.
\end{document}